\documentclass[a4paper,10pt,twoside]{cpc-hepnp}

\usepackage{multicol}
\usepackage{graphicx}
\usepackage{booktabs}
\usepackage{amssymb,bm,mathrsfs,bbm,amscd}
\usepackage[tbtags]{amsmath}
\usepackage{lastpage}

\begin{document}

\fancyhead[co]{\footnotesize X. Liu: Towards the decays of $\bar
N_X(1625)$ in the molecular picture}

\footnotetext[0]{Received ** ** 2009}

\title{Towards the decays of $\bar{N}_X(1625)$ in the molecular picture\thanks{Supported by National Natural Science
Foundation of China (10705001) }}

\author{%
      Xiang Liu$^{1;1)}$\email{liuxiang@teor.fis.uc.pt}%
}
\maketitle

\address{%
1~(School of Physical Science and Technology,
Lanzhou University, Lanzhou 730000, China)\\
}

\begin{abstract}

In this talk, we firstly overview the experimental status of
$\bar{N}_X(1625)$, which is an enhancement structure observed in
$K^{-}\bar{\Lambda}$ invariant mass spectrum of $J/\psi\to
pK^{-}\bar{\Lambda}$ process. Then we present the result of the
decay of $\bar{N}_X(1625)$ under the two molecular assumptions,
i.e. S-wave $\bar\Lambda K^-$ and S-wave $\bar{\Sigma}^0K^-$
molecular states. Several experimental suggestions for
$\bar{N}_X(1625)$ are proposed.

\end{abstract}



\begin{multicols}{2}

\section{Introduction}

$J/\psi$ decay is an ideal platform for studying the excited
baryons and hyperons. With the collected data, the BES experiment
has carried out a series of investigations of hadron spectroscopy.
Among the new observations of the hadron states, $\bar{N}_X(1625)$
is an enhancement near $K^{-}\bar{\Lambda}$ threshold, which was
only reported in several conference
proceedings\cite{HXYANG,JIN,SHEN} under the investigation of
$K^{-}\bar{\Lambda}$ invariant mass spectrum in $J/\psi\to
pK^{-}\bar{\Lambda}$ process. The rough measurement results about
the mass and the width of $\bar{N}_X(1625)$ are $m=1500\sim1650$
MeV and $\Gamma=70\sim110$ MeV, respectively. The experiment also
indicates that the spin-parity favors $\frac{1}{2}^-$ for
${N}_X(1625)$, which denotes the antiparticle of
$\bar{N}_X(1625)$\cite{SHEN}. The $pK^- \bar \Lambda$ Dalitz plot
and $K^- \bar \Lambda$ invariant mass spectrum are shown in figs.
\ref{fig1} and \ref{fig2}. $N_X(1625)$ enhancement structure was
not observed in $\gamma p\to K^+\Lambda$ process at
SAPHIR\cite{saphir}.
\begin{center}
\includegraphics[width=5cm]{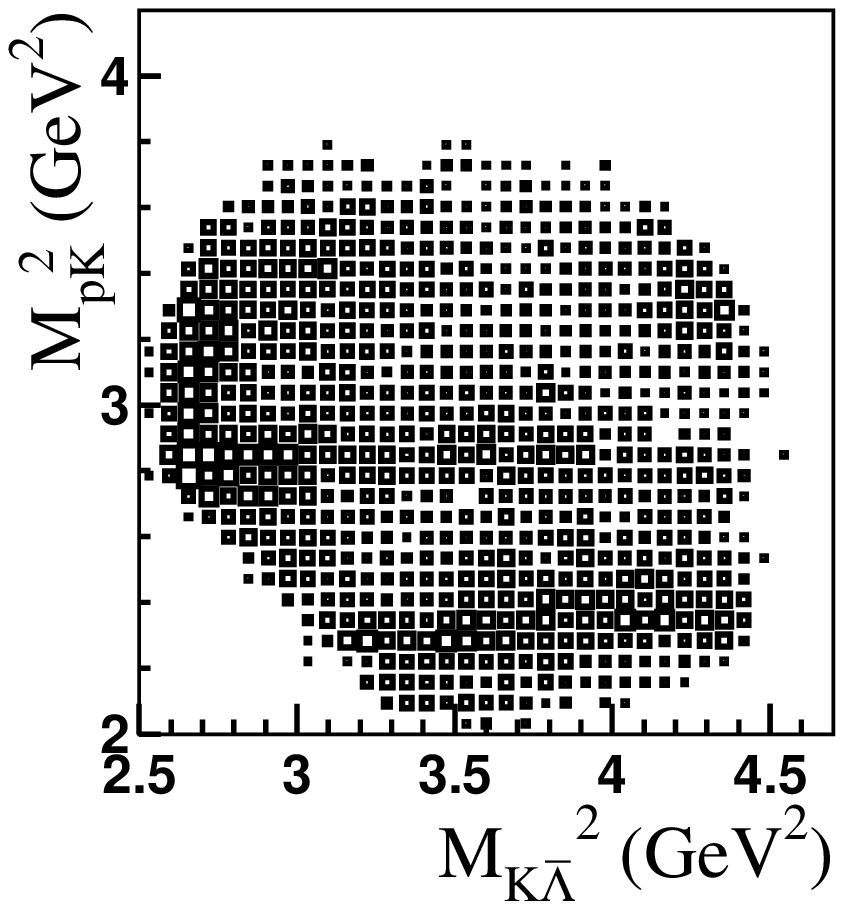}
\figcaption{\label{fig1}The Dalitz plot of $J/\psi\to pK^- \bar
\Lambda$ in Ref.~\citep{SHEN}.}
\end{center}
\begin{center}
\includegraphics[width=8cm]{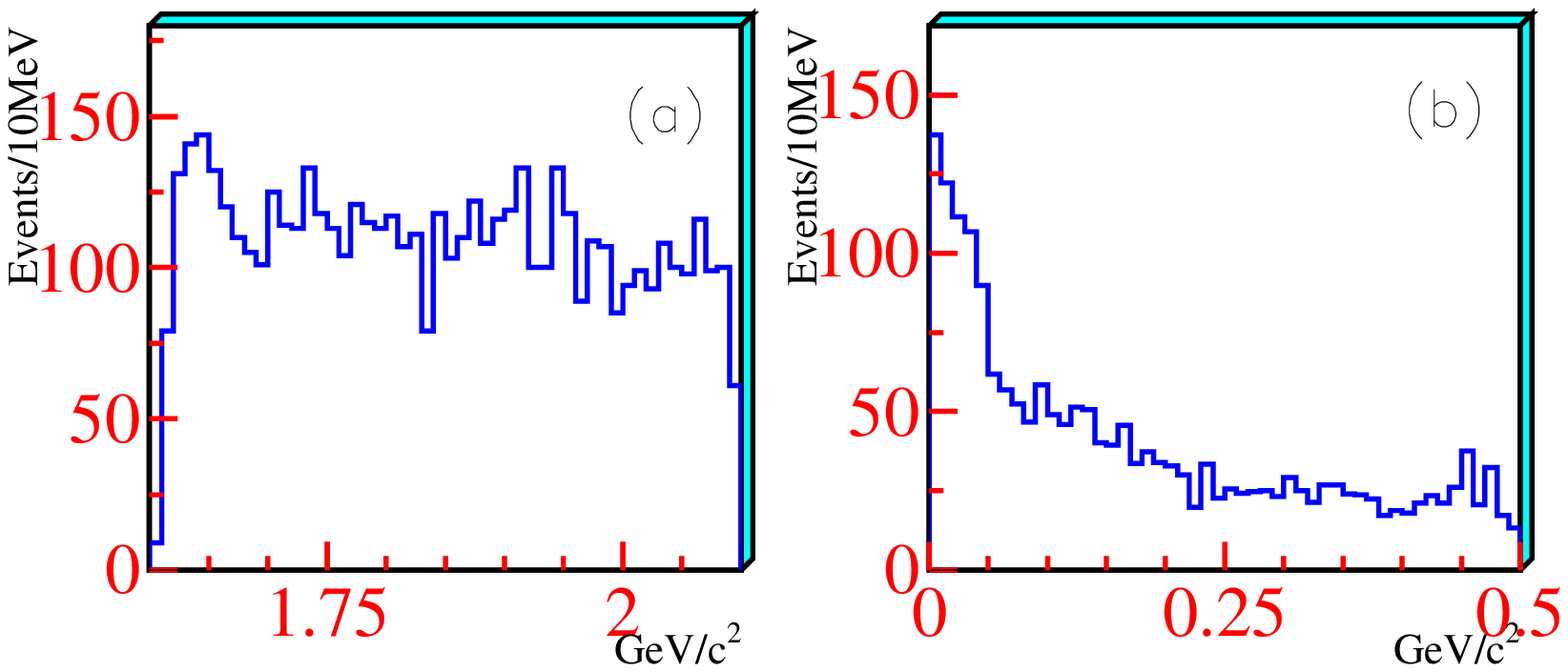}
\figcaption{\label{fig2} The invariant mass spectrum (a) $M_{K^-
\bar \Lambda}$ from $J/\psi \to pK^- \bar \Lambda$ and (b) the
$M_{K^- \bar \Lambda}-M_{K^-}-M_{\bar \Lambda}$ after the
efficiency and phase space correction from Ref.~\citep{SHEN}.}
\end{center}

At Hadron 07 conference, the BES Collaboration reported the
preliminary new experiment result of $\bar N_{X}(1625)$. Its mass
and width are well determined as\cite{BES-N(1625)}
$$m=1625^{+5+13}_{-7-23}\;\;\mathrm{MeV},\;\; \Gamma=43^{+10+28}_{-7\;\;-11}\;\;\mathrm{MeV}$$ respectively.
The production rate of $\bar{N}_X(1625)$ is
\begin{eqnarray*}
&&B[J/\psi\to p \bar{N}_{X}(1625)]\cdot B[\bar{N}_{X}(1625)\to
K^{-}\bar{\Lambda}]\\&&=(9.14^{+1.30+4.24}_{-1.25 -8.28})\times
10^{-5}.
\end{eqnarray*}
These more accurate experimental information of $\bar N_X(1625)$
provides us good chance to study the nature of $\bar N_X(1625)$.

If $\bar N_{X}(1625)$ is a regular baryon, the branching ratio of
$J/\psi\to p\bar{N}_{X}(1625)$ should be comparable with that of
$J/\psi\to p\bar{p}$ considering the branching ratio $B(J/\psi\to
p\bar{p})=2.17\times 10^{-3}$\cite{PDG}. Thus, we can obtain
$B[\bar N_{X}(1625)\to \bar{\Lambda}K^{-}]\sim 10\%$, which
indicates that there exists the strong coupling between
$\bar{N}_{X}(1625)$ and $K^{-}\bar{\Lambda}$.

This peculiar property of $\bar N_{X}(1625)$ inspires our interest
in exploring its structure, especially in its exotic component. In
Ref.~\citep{xiangliu-1625}, we calculated the possible decay modes
of $\bar{N}_X(1625)$ in the two different assumptions of the
molecular states, i.e. $\bar{\Lambda}-K^-$ and
$\bar{\Sigma}^{0}-K^-$. In the following, we will present the
details of the calculation and the numerical result.

\section{The decays under the assumptions of
$\bar{\Lambda}-K^-$ and $\bar{\Sigma}^0-K^-$ molecular states}

Since the mass of $\bar{N}_X(1625)$ is above the threshold of
$\bar{\Lambda}$ and $K^-$ under the assumptions of
$\bar{\Lambda}-K^-$ molecular state, thus $\bar{N}_X(1625)$ can
directly decay into $\bar{\Lambda}+K^-$ (Fig. \ref{decay} (a)),
which is depicted by the decay amplitude
\begin{eqnarray}
\mathcal{M}[\bar{N}_X(1625)\to
\bar{\Lambda}+K^-]=i\mathcal{G}\bar{v}_{N}\gamma_{5}{v}_{\bar{\Lambda}}.
\end{eqnarray}
Here $\mathcal{G}$ denotes the coupling constant between
$\bar{N}_X(1625)$ and $\bar{\Lambda}K^-$. $v_{\bar{\Lambda}}$ and
$v_{N}$ are the spinors.

\begin{center}
\includegraphics[width=7cm]{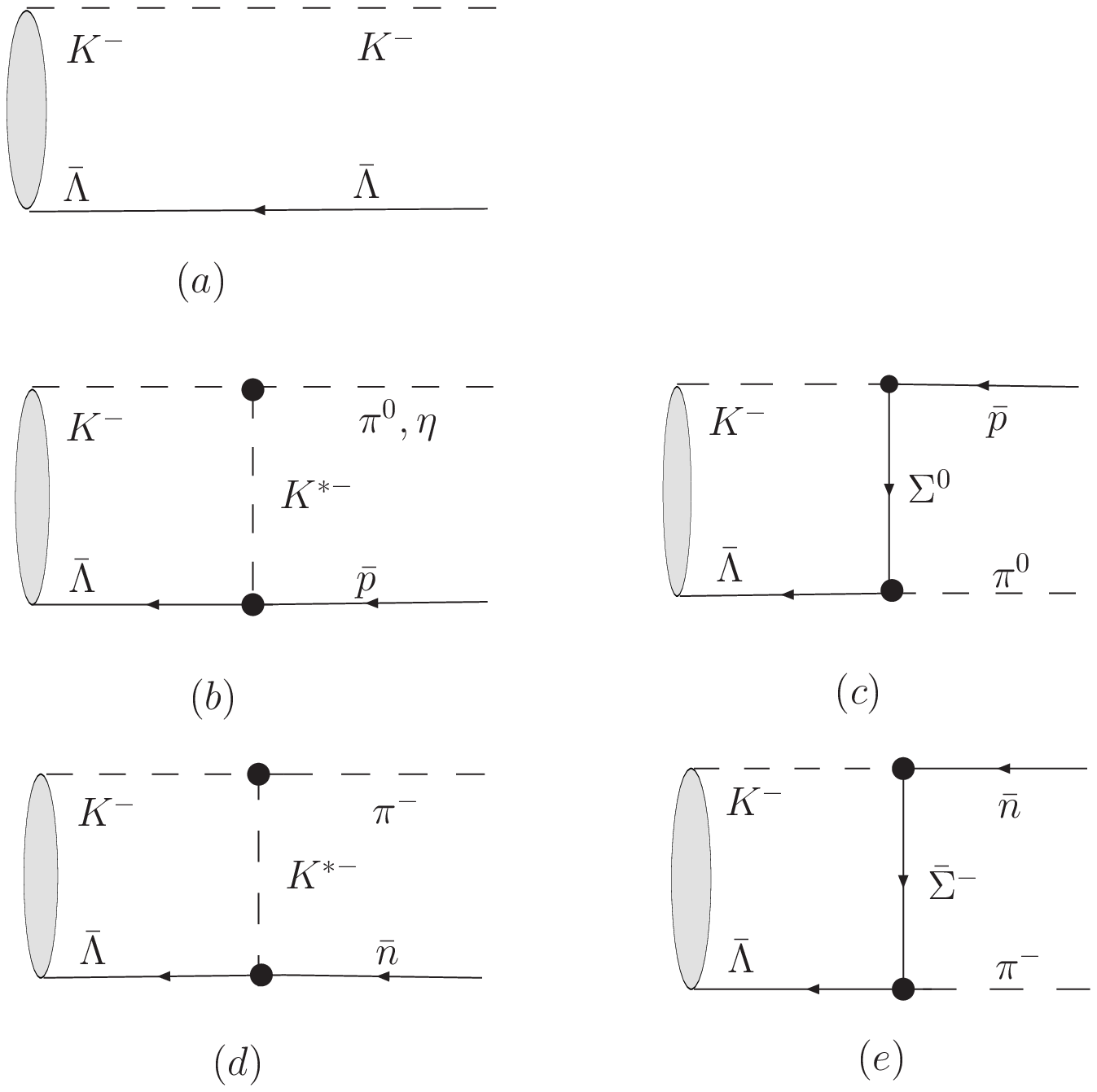}
\figcaption{\label{decay}The decay modes if $\bar{N}_X(1625)$ is
$\bar{\Lambda}-K^-$ molecular state.}
\end{center}

In the rescattering mechanism, the subordinate decays
$\bar{N}_X(1625)\to\pi^{0}\bar{p},\,\eta\bar{p},\,\pi^{-}\bar{n}$
occur, which are depicted in Fig. \ref{decay} (c)-(e). The
effective Lagrangians relevant to the calculation
are\cite{swart,lagrangian}:
\begin{eqnarray}
\mathcal{L}_{\mathcal{PPV}}&=&-ig_{_{\mathcal{PPV}}}Tr\big([\mathcal{P},\partial_{\mu}\mathcal{P}]\mathcal{V}^{\mu}\big),\label{ppv}\\
\mathcal{L}_{\mathcal{BBP}}&=&F_{P}Tr\big(\mathcal{P}[\mathcal{B},\bar{\mathcal{B}}]\big)\gamma_{5}+D_{P}Tr\big(\mathcal{P}\{\mathcal{B},\bar{\mathcal{B}}\}
\big)\gamma_{5},\label{BBP}\\
\mathcal{L}_{\mathcal{BBV}}&=&
F_{V}Tr\big(\mathcal{V}^{\mu}[\mathcal{B},\bar{\mathcal{B}}]\big)\gamma_{\mu}
+D_{V}Tr\big(\mathcal{V}^{\mu}\{\mathcal{B},\bar{\mathcal{B}}\}\big)\gamma_{\mu},\nonumber\label{BBV}\\
\end{eqnarray}
where $\bar{B}$ is the Hermitian conjugate of $B$. $\mathcal{P}$,
$\mathcal{V}$ and $B$ respectively denote the octet pseudoscalar
meson, the nonet vector meson and the baryon matrices. $F_P$ and
$D_P$ in eq. (\ref{BBP}) and $F_V$ and $D_V$ in eq. (\ref{BBV})
satisfy the relations $F_{P}/D_{P}=0.6$ \cite{F/D} and
$F_{V}/(F_{V}+D_{V})=1$\cite{thesis}. In the limit of SU(3)
symmetry, by $g_{NN\pi}=13.5$ and
$g_{NN\rho}=3.25$\cite{coupling}, one obtains the meson-baryon
coupling constants relevant to our calculation: $g_{PPV}=6.1$,
$F_P=13.5$, $D_P=0$, $F_V=1.2$, $D_V=2.0$.

Since the intermediate states $\bar{\Lambda}$ and $K^{-}$ in Fig.
\ref{decay} (b)-(d) are on-shell, one writes out the general
amplitude expression corresponding to Fig. \ref{decay} (b) and (d)
by Cutkosky cutting rules
\begin{eqnarray}
\mathcal{M}_{1}^{(\mathcal{A}_1,\mathcal{C}_1)}&=&\frac{1}{2}\int\frac{d^3
p_{1}}{(2\pi)^{3}2E_{1}}\frac{d^3
p_{2}}{(2\pi)^{3}2E_{2}}\nonumber\\
&&\times(2\pi)^{4}\delta^{4}(M_{N}-p_{1}-p_{2})[i\mathcal{G}\bar{v}_{N}\gamma_{5}{v}_{\bar{\Lambda}}]
\nonumber\\
&&\times
[ig_{1}\bar{v}_{{\bar{\Lambda}}}\gamma_{\mu}v_{{\mathcal{A}_1}}][ig_{2}(p_{1}+p_{3})_{\nu}]\frac{i}{q^{2}-M_{\mathcal{C}_1}^{2}}
\nonumber\\
&&\times\Big[-g^{\mu\nu}+\frac{q^{\mu}q^{\nu}}{M_{\mathcal{C}_1}^2}\Big]\mathcal{F}^{2}(M_{\mathcal{C}_1},q^2).
\end{eqnarray}
For Fig. \ref{decay} (c) and (e), the general amplitude expression
is
\begin{eqnarray}
\mathcal{M}_{1}^{(\mathcal{A}_2,\mathcal{C}_2)}&=&\frac{1}{2}\int\frac{d^3
p_{1}}{(2\pi)^{3}2E_{1}}\frac{d^3
p_{2}}{(2\pi)^{3}2E_{2}}\nonumber\\
&&\times(2\pi)^{4}\delta^{4}(M_{N}-p_{1}-p_{2})[i\mathcal{G}\bar{v}_{N}\gamma_{5}v_{\bar{\Lambda}}]
\nonumber\\
&&\times [ig_{2}'\bar{v}_{\bar{\Lambda}}\gamma_{5}]
\frac{i(q\!\!\!\slash+M_{\mathcal{C}_2})}{q^{2}-M_{\mathcal{C}_2}^{2}}[ig_{1}'\gamma_{5}{v}_{\mathcal{A}_2}]
\nonumber\\
&&\times\mathcal{F}^{2}(M_{\mathcal{C}_2},q^2).
\end{eqnarray}
In the above expressions, $\mathcal{C}_{i}$ and $\mathcal{A}_{i}$
denote the exchanged particle and the final state baryon,
respectively. $p_{1}$ and $p_{2}$ are respectively the four
momenta of $K^-$ and $\bar{\Lambda}$. $\mathcal{F}^{2}(m_{i},q^2)$
denotes the form factor which compensates the off-shell effects of
the hadrons at the vertices. In this work, one takes
$\mathcal{F}^{2}(m_{i},q^2)$ as the monopole form\cite{FF,HY-Chen}
$\mathcal{F}^{2}(m_{i},q^2)=\bigg(\frac{\xi^{2}-m_{i}^2
}{\xi^{2}-q^{2}}\bigg)^2$, which plays the role to cut off the end
effect. Phenomenological parameter $\xi$ is parameterized as
$\xi=m_{i}+\alpha \Lambda_{QCD}$, where $m_{i}$ denotes the mass
of exchanged meson\cite{HY-Chen} and $\Lambda_{QCD}=220$ MeV.
$\alpha$ is a phenomenological parameter and is of order unity.

In the $\bar{\Sigma}^0-K^-$ molecular picture, $\bar{N}_{X}(1625)$
does not decay into $\bar{\Sigma}^0$ and $K^-$ because of having
not enough phase space. However, decay $\bar{N}_X(1625)\to
\bar{\Lambda}+K^-$ occurs by the isospin violation effect, which
results in the mixing of $\Sigma^{0}$ with $\Lambda$\cite{s.l.
zhu} (see Fig. \ref{decay1} (a)). By the Lagrangian
$$\mathcal{L}_{\mbox{mixing}}=g_{\mbox{mixing}}(\bar{\psi}_{{\Sigma}^0}
\psi_{{\Lambda}}+\bar{\psi}_{{\Lambda}}{\psi}_{{\Sigma}^0})$$ with
the coupling constant $g_{\mbox{mixing}}=0.5\pm 0.1$ MeV
determined by QCD sum rule \cite{s.l. zhu}, one writes out the
decay amplitude
\begin{eqnarray}
&&\mathcal{M}[\bar{N}_X(1625)\to
\bar{\Sigma}^{0}+K^-]\nonumber\\&&\quad\quad=\mathcal{G}\;g_{\mbox{mixing}}\;
\bar{v}_{N}\gamma_{5}\frac{i}{p\!\!\!\slash-M_{\Lambda}}v_{\bar{\Lambda}},
\end{eqnarray}
where $p$ and $M_{\Lambda}$ are the four momentum and the mass of
$\bar{\Lambda}$, respectively.

\begin{center}
\includegraphics[width=7cm]{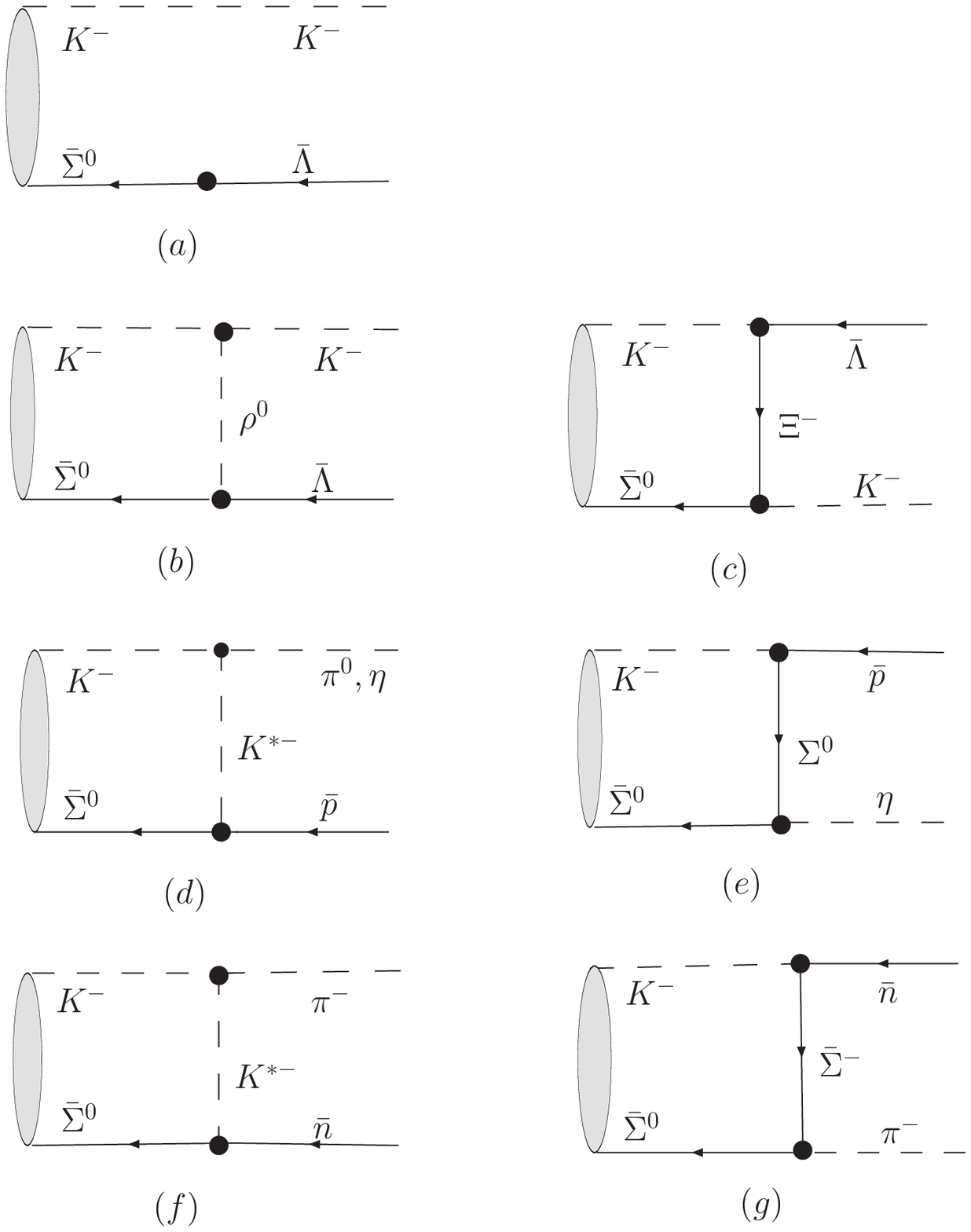}
\figcaption{\label{decay1}The decay modes if $\bar{N}_X(1625)$ is
$\bar{\Sigma}^0-K^-$ molecular state.}
\end{center}

For $\bar{\Sigma}^0-K^-$ molecular state assumption,
$\bar{N}_X(1625)$ still can decay into
$\pi^{0}\bar{p},\,\eta\bar{p},\,\pi^{-}\bar{n}$, which are
described in Fig. \ref{decay1} (b)-(g). The general expression of
Fig. \ref{decay1} (b), (d), (f) is expressed as
\begin{eqnarray}
\mathcal{M}_{3}^{(\mathcal{A}_3,\mathcal{C}_3)}&=&\int\frac{d^4
q}{(2\pi)^{4}}[i\mathcal{G}\bar{v}_{N}\gamma_{5}]
\frac{i}{-\not{p}_{2}-M_{\bar{\Sigma}^{0}}}[ig_{3}\gamma_{\mu}{v}_{{\mathcal{A}_3}}]
 \nonumber\\&&
\times[ig_{4}(p_{1}+p_{3})_{\nu}]\frac{-ig^{\mu\nu}}{q^{2}-M_{\mathcal{C}_3}^{2}}
\frac{i}{p_{1}^{2}-M_{K}^2}\nonumber\\&&
\times\mathcal{F}^{2}(M_{\mathcal{C}_3},q^2),\label{hadron-loop-1}
\end{eqnarray}
for Fig. \ref{decay1} (c), (e), (g) the general amplitude
expression reads as
\begin{eqnarray}
\mathcal{M}_{4}^{(\mathcal{A}_4,\mathcal{C}_4)}&=&\int\frac{d^4
q}{(2\pi)^{4}}[i\mathcal{G}\bar{v}_{N}\gamma_{5}]\frac{i(\not{p}_{2}-M_{\bar{\Sigma}^0})}{-p_{2}^{2}-M_{\bar{\Sigma}^0}^{2}}[ig_{4}'\gamma_{5}]
\nonumber\\
&&\times
\frac{i(q\!\!\!\slash+M_{\mathcal{C}_4})}{q^{2}-M_{\mathcal{C}_4}^{2}} [ig_{3}'\gamma_{5}{v}_{\mathcal{A}_4}]\nonumber\\
&&\times\frac{i}{p_{1}^{2}-M_{K}^{2}}
\mathcal{F}^{2}(M_{\mathcal{C}_4},q^2),\label{hadron-loop-2}
\end{eqnarray}
where $p_{1}$ and $p_2$ denote the four momenta carried by $K^-$
and $\bar{\Sigma}^0$, respectively. $q=p_1-p_3=p_4-p_2$. For the
decays depicted in Fig. \ref{decay1} (b)-(g), $\bar{\Sigma}^0$ and
$K^-$ are off-shell. The form factor may provide a convergent
behavior for the triangle loop integration, which is very similar
to the case of the Pauli-Villas renormalization
scheme\cite{Izukson,peskin,liuxiang}.

\section{Numerical result}

In Figs. \ref{ratio-1} and \ref{ratio-2}, we show the ratios of
the decay widths of
$\bar{N}_X(1625)\to\pi^0\bar{p},\eta\bar{p},\pi^{-}\bar{n}$ to the
decay width of $\bar N_X(1625)\to \bar{\Lambda} K^-$ under the
assumptions of $\bar{\Lambda}-K^-$ and $\bar{\Sigma}^0-K^-$
molecular states when taking $\alpha=1\sim 3$. Fig. \ref{ratio-1}
and Fig. \ref{ratio-2} illustrate that these ratios do not
strongly depend on the $\alpha$. One further obtains the typical
values of these ratios taking $\alpha=1.5$, which are listed in
Table \ref{ratio-all}. Combining these ratios shown in Figs.
\ref{ratio-1} and \ref{ratio-2} with the branching ratio
$B[J/\psi\to p \bar{N}_{X}(1625)]\cdot B[\bar{N}_{X}(1625)\to
K^{-}\bar{\Lambda}]=(9.14^{+1.30+4.24}_{-1.25 -8.28})\times
10^{-5}$ given by BES \cite{BES-N(1625)}, one estimates the
branching ratio of the subordinate decays of $J/\psi\to p
\bar{N}_{X}(1625)\to
p(\pi^0\bar{p}),\,p(\eta\bar{p}),\,p(\pi^-\bar{n})$, which are
shown in Table. \ref{numerical}.

\begin{center}
\includegraphics[width=9cm]{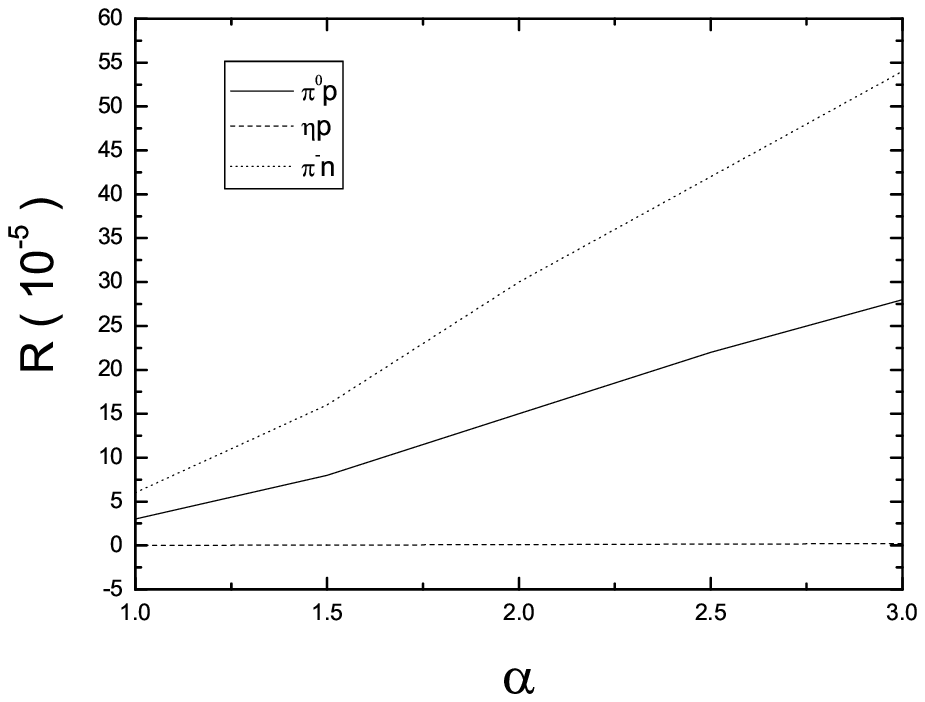}
\figcaption{\label{ratio-1}The ratios of $\bar
N_X(1625)\to\pi^0\bar{p},\eta\bar{p},\pi^{-}\bar{n}$ decay widths
to $\bar{N}_X(1625)\to \bar{\Lambda} K^-$ decay width under the
assumption of $\bar{\Lambda}-K^-$ molecular state.}
\end{center}

\begin{center}
\includegraphics[width=9cm]{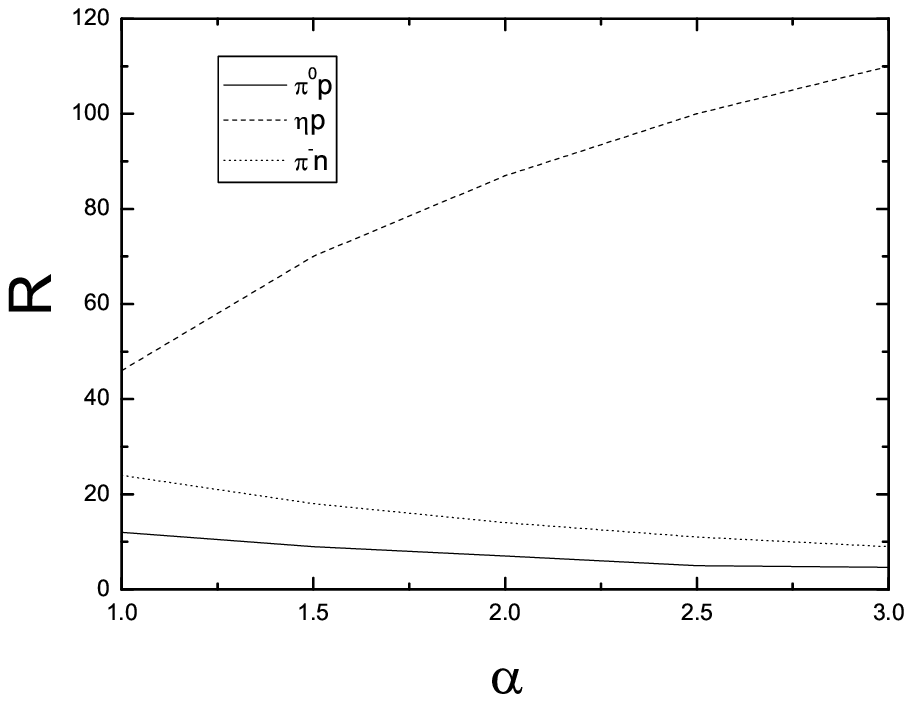}
\figcaption{\label{ratio-2}The ratios of $\bar
N_X(1625)\to\pi^0\bar{p},\eta\bar{p},\pi^{-}\bar{n}$ decay widths
to $\bar N_X(1625)\to \bar{\Lambda} K^-$ decay width in
$\bar{\Sigma}^0-K^-$ molecular state picture.}
\end{center}

\begin{center}
\tabcaption{The ratios of the decay widths of
$\bar{N}_X(1625)\to\pi^0\bar{p},\eta\bar{p},\pi^{-}\bar{n}$ to the
decay width of $\bar{N}_X(1625)\to \bar{\Lambda} K^-$ in different
molecular assumptions with $\alpha=1.5$.\label{ratio-all}}
\footnotesize
\begin{tabular*}{80mm}{c@{\extracolsep{\fill}}ccc}
\toprule  &
$\frac{\Gamma(\pi^{0}\bar{p})}{\Gamma(K^{-}\bar{\Lambda})}$&$\frac{\Gamma(\eta\bar{p})}{\Gamma(K^{-}\bar{\Lambda})}
$&$\frac{\Gamma(\pi^{-}\bar{n})}{\Gamma(K^{-}\bar{\Lambda})}$ \\
\hline $\bar{\Lambda}-K^-$&$1\times
10^{-4}$&$5\times10^{-7}$&$2\times
10^{-4}$ \\
$\bar{\Sigma}^0-K^-$&9&$70$&18\\
\bottomrule
\end{tabular*}
\end{center}
\end{multicols}

\begin{center}
\tabcaption{The branching ratios of $J/\psi\to p
\bar{N}_{X}(1625)\to
p(\pi^0\bar{p}),\,p(\eta\bar{p}),\,p(\pi^-\bar{n})$ in two
different molecular state pictures for $\bar{N}_X(1625)$.
\label{numerical}} \footnotesize
\begin{tabular*}{100mm}{c@{\extracolsep{\fill}}ccc}
\toprule & $\bar{\Lambda}-K^{-}$ system &$\bar{\Sigma}^{0}-K^{-}$ system \\
\hline$J/\psi\to p \bar N_X(1625)\to p(\pi^{0}\bar{p})$& $1 \times 10^{-8} \sim 3 \times 10^{-8} $& $\sim 1 \times 10^{-3} $ \\
$J/\psi\to p \bar N_X(1625)\to p(\eta\bar{p})$   & $4 \times 10^{-11}\sim 2 \times 10^{-10}$& $\sim 7 \times 10^{-3} $ \\
$J/\psi\to p \bar N_X(1625)\to p(\pi^{-}\bar{n})$& $2 \times 10^{-8} \sim 5 \times 10^{-8} $& $\sim 2 \times 10^{-3} $ \\
\bottomrule
\end{tabular*}
\end{center}

\vspace{0.5cm}
\begin{multicols}{2}

\section{Discussion and conclusion}

Assuming $\bar{N}_{X}(1625)$ as $\bar{\Lambda}-K^-$ molecular
state, $K^{-}\bar{\Lambda}$ is the dominant decay mode of
$\bar{N}_X(1625)$. The branching ratio of $\bar N_X(1625)\to
K^{-}\bar{\Lambda}$ is far larger than the branching ratios of
$\bar{N}_X(1625)\to\pi^{0}\bar{p},\eta\bar{p},\pi^{-}\bar{n}$,
which can explain why $\bar N_X(1625)$ was firstly observed in the
mass spectrum of $K^{-}\bar{\Lambda}$. And we notice that the
smallest measurable branching ratio for $J/\psi$ decay listed in
the Particle Data Book\cite{PDG} is about $10^{-5}$. Thus, it is
difficult to measure $J/\psi\to p\bar N_X(1625)\to
p(\pi^{0}\bar{p}),\,p(\eta\bar{p}),\,p(\pi^{-}\bar{n})$ in further
experiments.

Under the assumption of S-wave $\bar{\Sigma}^{0}-K^-$ molecular
state for $\bar N_{X}(1625)$, $\bar N_X(1625)$ can not decay to
$\bar{\Sigma}^{0}K^-$ due to having not enough phase space. The
$\Lambda-\Sigma^0$ mixing mechanism and final state interaction
effect result in the decay $\bar N_X(1625)\to\bar{\Lambda}K^-$.
The branching ratio of $\bar N_X(1625)\to\bar{\Lambda}K^-$ is
about one or two order smaller than that of $\bar N_X(1625)\to
\pi^{0}\bar{p}, \eta\bar{p}, \pi^{-}\bar{n}$. The sum of the
branching ratios of $\bar N_X(1625)\to \pi^{0}\bar{p},
\eta\bar{p}, \pi^{-}\bar{n}$ listed in Table \ref{numerical} is
about $10^{-2}$. Such a large branching ratio is unreasonable for
$J/\psi$ decay. The BES collaboration has already studied
$J/\psi\to p\pi^-\bar{n}$ in Ref. \citep{peta-2} and $J/\psi\to
p(\eta\bar{p})$ in Ref. \citep{peta-3}. The branching ratios
respectively corresponding to $J/\psi\to p\pi^-\bar{n}$ and
$J/\psi\to p\eta\bar{p}$ are $2.4\times 10^{-3}$ and $2.1\times
10^{-3}$\cite{peta-2,peta-3}. Although these experimental values
are comparable with our numerical result of the corresponding
channel, the former experiments did not find the structure
consistent with $\bar{N_{X}}(1625)$, which seems to show that the
evidence against S-wave $\bar{\Sigma}^0-K^-$ molecular picture is
gradually accumulating\cite{xiangliu-1625}.

As indicated in Ref. \citep{BES-N(1625)}, there exists very strong
coupling between $\bar N_X(1625)$ and $\bar{\Lambda}K^{-}$. At
present other decay modes of $\bar N_X(1625)$ are still missing
\cite{BES-N(1625)}. Thus the assumption of S-wave
$\bar{\Lambda}-K^-$ molecular state is more favorable than that of
S-wave $\bar{\Sigma}^0-K^-$ molecular state for $\bar N_X(1625)$.
The result of Ref. \citep{ZHANG}, which is from the calculation
within the framework of the chiral $SU(3)$ quark model by solving
a resonating group method (RGM) equation, indicates that the
$\Lambda K$ system is unbound. Whether there exists the S-wave
$\bar{\Lambda}-K^-$ molecular state is still an open issue. The
dynamics study of S-wave $\bar{\Lambda}-K^-$ system by other
phenomenological models is encouraged.

If it is problematic to explain $ \bar{N}_X(1625)$ as the pure
molecular state structure, we have to again ask what is the
underlying structure of $\bar{N}_X(1625)$. We notice that there
exist two well established states $N^*(1535)$ and $N^*(1650)$ with
$J^{P}=1/2^-$ nearby the mass of ${N}_{X}(1625)$. In
PDG\cite{PDG}, the branching ratio of $N^{*}(1650)\to K\Lambda $
is about $3\sim 11\%$. The authors of Ref. \citep{LIU} indicated
that $N^*(1535)$ should have large $s\bar{s}$ component in its
wave function which shows the large $N^*(1535)K\Lambda$ coupling.
$N^*(1535)$ and $N^*(1650)$ can strongly couple to $K\Lambda$.
Thus, whether $N_X(1625)$ enhancement is related to $N^*(1535)$
and $N^*(1650)$ is also an interesting topic.

Finally, we want to propose several suggestions for future
experiment:

\begin{itemize}
{\item Until now, the experimental information of
$\bar{N}_X(1625)$ only appeared in the proceeding of
conference\cite{HXYANG,JIN,SHEN,BES-N(1625)}. We are expecting the
formal publication of this enhancement, which will be helpful to
stimulate more experimentalists and theorists to pay attention to
this issue. }

{\item Searching for $\bar N_X(1625)\to \pi^{0}\bar{p},
\eta\bar{p}, \pi^{-}\bar{n}$ modes in future experiment can shed
light on the nature of $\bar N_X(1625)$. We urge our experimental
colleague carefully analyze $J/\psi\to p\pi^-\bar{n}$ and
$J/\psi\to p\eta\bar{p}$ channel in further experiments,
especially in the forthcoming BESIII.}

{\item Confirming $\bar N_X(1625)$ by the other experiments is
encouraged. At present, Lanzhou CSR is a good platform to study
the baryon spectroscopy. Analyzing the invariant mass spectrum of
$K^+\Lambda$}, which comes from the $p\alpha$ reaction, will be an
important approach to investigate the $N_X(1625)$ enhancement
structure.

\end{itemize}

\acknowledgments{We thank the organizer of Workshop on the Physics
of Excited Nucleon- NSTAR2009 for providing us a good chance to
communicate the research work with each other. X.L also enjoys the
collaboration with Bo Zhang.}

\end{multicols}

\vspace{-2mm}
\centerline{\rule{80mm}{0.1pt}}
\vspace{2mm}

\begin{multicols}{2}

\end{multicols}

\clearpage

\end{document}